# A multi-component Fermi surface in the vortex state of an underdoped high-$T_c$ superconductor

Suchitra E. Sebastian[1], N. Harrison[2], E. Palm[3], T. P. Murphy[3], C. H. Mielke[2], Ruixing Liang[4,5], D. A. Bonn[4,5], W. N. Hardy[4,5] & G. G. Lonzarich[1]

1. Cavendish Laboratory, Cambridge University, JJ Thomson Avenue, Cambridge CB3 0HE, U.K.

2. NHMFL, Los Alamos National Laboratory, MS E536, Los Alamos, NM 87545, U.S.A.

3. National High Magnetic Field Laboratory (NHMFL), Florida State University, Tallahassee, Florida 32306, U.S.A.

4. Department of Physics and Astronomy, University of British Columbia, Vancouver V6T 1Z4, Canada.

5. Canadian Institute for Advanced Research, Toronto M5G 1Z8, Canada.

**In order to understand the origin of superconductivity, it is crucial to ascertain the nature and origin of the primary carriers available to participate in pairing[1-6]. Recent quantum oscillation experiments on high $T_c$ cuprate superconductors[7-10] have revealed the existence of a Fermi surface akin to normal metals, comprising fermionic carriers that undergo orbital quantization[11]. However, the unexpectedly small size of the observed carrier pocket leaves open a variety of possibilities as to the existence or form of any underlying magnetic order, and its relation to $d$-wave superconductivity[12-15]. Here we present quantum oscillations in the magnetisation (the de Haas-van Alphen or dHvA effect) observed in superconducting YBa$_2$Cu$_3$O$_{6.51}$ that reveal more than one carrier pocket. In particular, we find evidence for the existence of a much larger pocket of heavier mass carriers playing**

**a thermodynamically dominant role in this hole-doped superconductor. Importantly, characteristics of the multiple pockets within this more complete Fermi surface impose constraints on the wavevector of any underlying order and the location of the carriers in momentum space. These constraints enable us to construct a possible density-wave scenario with spiral or related modulated magnetic order, consistent with experimental observations.**

$YBa_2Cu_3O_{6+x}$ belongs to the family of quasi-two dimensional (Q2D) cuprate superconductors that can exhibit surprisingly high critical temperatures compared to other layered superconductors[16]. Until recently[7], closed Fermi surface sheets in these high $T_c$ superconductors had posed a considerable challenge to observe, requiring very strong pulsed magnetic fields. In the present experiment we utilise the magnetic torque technique[17-18] in continuous magnetic fields, which is particularly favourable given the Q2D structure of these materials, enabling us to measure dHvA effect oscillations under carefully controlled experimental conditions. Figure 1 shows an example of such a measurement on a single crystal of underdoped $YBa_2Cu_3O_{6.51}$ (a UBC-grown sample on which quantum oscillations are reported in ref. 7).

The key finding of the present study is the observation of a new oscillatory component (β) of higher frequency $F_β = 1654 ± 40$ T but almost 30 times smaller in amplitude than the dominant component of frequency first observed in ref. 7 (denoted as $F_α$ in this work – Fig. 2). The comparably weak amplitude of the second pocket (β) belies its greater thermodynamic significance— an important consideration in the determination of the carriers most relevant in pairing in this Q2D material. The effective mass of the β pocket $m_β = 3.8 ± 0.4\ m_e$ is approximately twice that of the α pocket – obtained from a temperature-dependent fit to the Lifshitz-Kosevich theory[11] as shown in Fig. 3. This new pocket also contains $3.29 ± 0.15$ times as many carriers as the α pocket, as determined from the Onsager expression[11] $A_k = 2πeF/\hbar$ relating the cross-sectional area to the dHvA frequency.

Were the α and β pockets to constitute the entire Fermi surface, their total carrier density is expected to correspond to the nominal hole doping value $p_{nom} ≈ 0.1$ (estimated in ref. 20). The effective carrier density on assuming Kramer's degeneracy is estimated by the summation $p_{eff} = 2(n_α A_{kα} + n_β A_{kβ})/A_{BZ}$ over pocket $k$-space areas, where $A_{BZ}$ is the area of the unreconstructed Brillouin zone and $n$ refers to the number of instances each occurs in the reconstructed Brillouin zone (negative for electrons and positive for holes). We find that a scenario where $n_α = -1$ and $n_β = 1$ (such that $p_{eff} = 0.082 ± 0.003$) yields reasonable consistency with $p_{nom}$, suggesting that the α-pocket is an electron pocket, and the β-pocket a hole pocket.

Given the Q2D electronic structure, the total pocket contribution to the linear temperature coefficient of the specific heat is given by the summation $γ_m = γ'(|n_α|m_α^* + |n_β|m_β^*)/m_e$, where $γ' = 1.46$ mJmol$^{-1}$K$^{-2}$ [ref. 8], provided once again that Kramer's degeneracy holds. A single pocket of each type (i.e. $|n_α| = 1$ and $|n_β| = 1$) would yield $γ_m ≈ 8.2 ± 0.6$ mJmol$^{-1}$K$^{-2}$, comparable in value to $γ_C ≈ 10$ mJmol$^{-1}$K$^{-2}$ measured in heat capacity experiments in the normal state above $T_c$ [ref. 19] (details in supplementary information).

If we associate Fermi surface reconstruction with the observed pockets, our experimental findings place constraints on the ordering wavevector **Q**. We start by making the simplifying single-layer assumption that **Q** averages out the effects of bilayer splitting within the YBa$_2$Cu$_3$O$_{6.51}$ unit cell by coupling bonding and anti-bonding states. Reconstruction of the Fermi surface shown in Fig. 4a by the commensurate ordering wavevector **Q** = $(\pi,\pi)$[14,15,21] (as seen in the Néel phase for $p_{nom}$ ~< 0.05, see ref. 16) leads to two identical hole pockets at equivalent **k** = $(\pi/2,\pi/2)$ points for each electron pocket at **k** = $(\pi,0)$ in the reconstructed Brillouin zone (shown in Fig. 4b). Under these circumstances where $n_\alpha = -1$ and $n_\beta = 2$, the estimated value of $p_{eff} = 2(n_\alpha A_{k\alpha} + n_\beta A_{k\beta})/A_{BZ} \approx 0.20 \pm 0.01$ would be more than twice the nominal doping value $p_{nom}, \approx 0.10$. The observed orbit sizes therefore appear too large to be compatible with a **Q** = $(\pi,\pi)$ scenario (Fig. 4b).

We therefore consider a scenario with either a differently commensurate order (similar to that considered in ref. 13) or truly incommensurate order, and show that this could lead to a hole pocket compatible in size with the observed β pocket. We consider the particular case of a single wave vector **Q** = $(\pi[1-2\delta],\pi)$, $(\pi[1+2\delta],\pi)$, $(\pi,\pi[1-2\delta])$ or $(\pi,\pi[1+2\delta])$ in YBa$_2$Cu$_3$O$_{6.51}$ where δ is taken to be 0.1 (see Fig. 4c). Such a wavevector corresponds to a helical or spiral density-wave scenario in which δ modulates the orientation of the staggered spins within the planes. This value of δ falls within the range seen in inelastic neutron scattering measurements for $p_{nom}$ ~ 0.1 that yield spin fluctuations at wavevectors **Q** = $(\pi[1\pm2\delta], \pi)$ [ref. 22], although long range order has not been observed at zero magnetic field in YBa$_2$Cu$_3$O$_{6.51}$ [ref. 23]. In such a scenario, the hole pockets at equivalent **k** = ($\pi/2,\pi/2$) points would be significantly asymmetric in size. In the limit of a sufficiently large ordering gap or $\delta$, only a single hole pocket would appear for each electron pocket (such that $n_\beta = 1$ instead of 2). In this case, the effective halving of the number of hole pockets with respect to the commensurate case results in a doubling of the hole pocket size, thereby reducing the density of states. Pocket sizes consistent with our experimental findings can arise from such a Fermi surface reconstruction (schematic in Fig. 4d), yielding a value of $p_{eff} = 0.082$ – reasonably close to $p_{nom}$, with a similar agreement for $\gamma_C$. Collinear structures, on the other hand (in which the amplitude of the spin is modulated by $\delta$), would involve multiple $\pm$**Q** vectors (as recently proposed for $\delta = 1/8$ in ref. 13) and in general lead to a hierarchy of gaps and smaller or no hole pockets[13,24]. In this case, prominent orbits like the $\beta$ orbit in Fig. 4d could become observable as a result of magnetic breakdown tunnelling overcoming all but the dominant gap in strong magnetic fields, as occurs in elemental Cr [ref. 24]. These scenarios discussed with a different commensurability or incommensurability may be equally applicable to all forms of density wave order[2,4,14,15].

The bilayer potential – the effect of which was considered small in the simplified single-layer scenario considered thus far – may, however, play an important role in determining a complete picture of the Fermi surface. A significant bilayer potential could lift the assumed degeneracy of either or both pockets, leading to dual frequencies for the affected pocket, one of which could even be completely depopulated (more likely for the electron pocket given its smaller size). It might then be possible that the observed $\alpha$ and $\beta$ frequencies each correspond to split frequencies for either or both electron and hole pockets.

We conclude with a brief discussion of further issues raised by this study. The first concerns the origin of the potential Fermi surface reconstruction in relation to superconductivity. One possibility is that modulated magnetic ordering competes with superconductivity, nucleating wherever superconductivity is suppressed, as in the vortex core regions. In the case of antiferromagnetic order, this behaviour may have been identified in heavy fermion compounds[26], and has been suggested by neutron scattering on $La_{2-x}Sr_xCuO_4$ (ref. 27) and muon spin rotation on $YBa_2Cu_3O_{6.5}$ (ref. 28). Suppression of antiferromagnetism by the onset of superconductivity may account for the loss of neutron scattering intensity at the incommensurate wavevectors at low energies and low temperatures observed in $YBa_2Cu_3O_{6.5}$ (ref. 23).

The second issue concerns the comparative stability of helical (or spiral)[25] versus collinear forms of order[3,13], which could affect the coexistence of superconductivity with magnetic order. Whereas the former favours a uniform distribution of doped-holes[25], the latter is conducive to formation of stripes[3]. In either scenario, magnetic field-induced Landau quantisation consistent with the observation of quantum oscillations could play an important role in stabilising long-range order or tuning **Q** in strong magnetic fields, as observed in other layered metals[29].

A third issue concerns the renormalisation of the cyclotron mass for the two Fermi surface pockets identified. A parameterisation of the unreconstructed band structure for $YBa_2Cu_3O_{6.51}$, such as that adopted in ref. 13, predicts pockets of similar unrenormalised band masses. The higher mass renormalisation of hole pockets suggested by our observations would, therefore, need to be taken into consideration within any explanation of superconductivity in this system.

Acknowledgements: This work was supported by the National Science Foundation, the Department of Energy (US), Florida State, the UK EPSRC, the Canadian Institute for Advanced Research, and NSERC. S.E.S acknowledges support from the Institute for Complex Adaptive Matter, COST, and Trinity College (Cambridge University). We acknowledge discussions with E. Abrahams, P. W. Anderson, E. Berg, A. Carrington, S. Chakravarty, L. P. Gor'kov, S. R. Julian, S. A. Kivelson, D. LeBoeuf, P. A. Lee, P. B. Littlewood, A. P. Mackenzie, A. Millis, M. R. Norman, D. Pines, C. Proust, T. M. Rice, S. Sachdev, L. Taillefer, and experimental assistance from G. Jones, J. H. Park, and S. Tozer.

Torque experiments were performed by SES, NH and GGL.

Reprints and permissions information is available at npg.nature.com/reprintsandpermissions

The authors declare no competing financial interests.

Correspondence and requests for materials should be addressed to S.E.S. (suchitra@phy.cam.ac.uk).

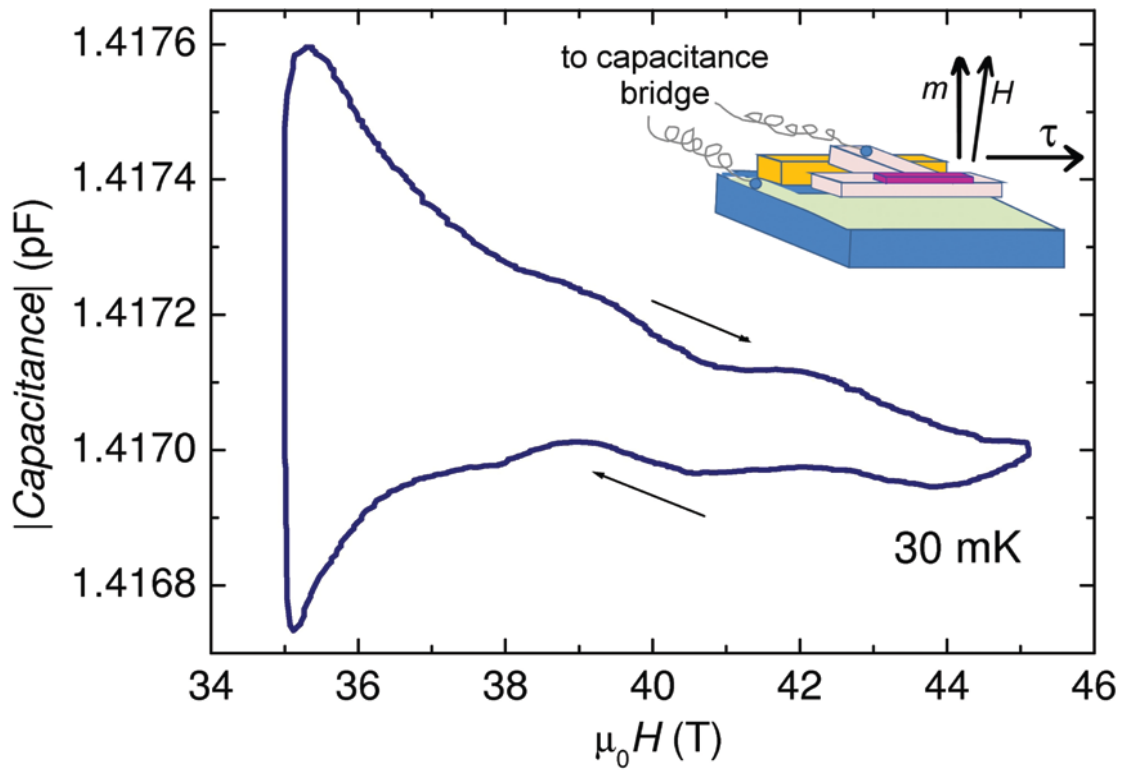

**Fig. 1. Experimental data**. Capacitance versus magnetic field obtained on sweeping the magnetic field up and down between 35 and 45 T at base temperature of the dilution refrigerator (~ 30 mK). The sample is placed on a cantilever whose deflection resulting from the magnetic torque $\tau = \mathbf{m} \times \mathbf{H}$ (due to a small inclination of the magnetic moment **m** with respect to the magnetic field **H**) is measured capacitively. The angle of inclination θ is minimised to within 3° of the crystalline *c*-axis, such that the angular deflection dθ of the cantilever due to quantum oscillations is of the order of 0° 0′ 1″. At zero magnetic field the capacitance is 1.4358 pF. The field-induced change consists of the superposition of quantum oscillatory components, dominated by the known component (α) of frequency $F_\alpha = 502 \pm 20$ T (this work – slightly lower

than refs. 7 & 10, see supplementary information) and cyclotron mass m* = 1.9 ± 0.1 $m_e$ [ref. 7], and irreversible superconducting components, establishing that quantum oscillations occur in a regime where some level of vortex pinning prevails (details in supplementary information). Our observation of dHvA oscillations within the mixed state with similar cyclotron orbit sizes and quasiparticle effective masses m* as those in the normal state[7,10] is reminiscent of other strongly type II superconductors[30].

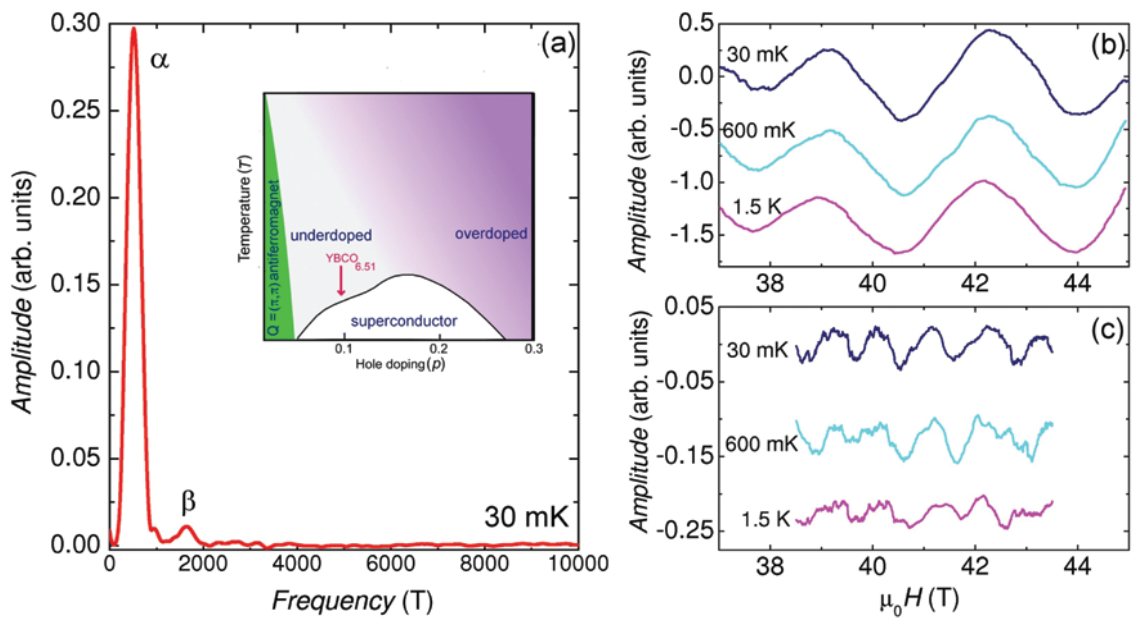

**Fig. 2. de Haas-van Alphen oscillations in $YBa_2Cu_3O_{6.51}$. a**, Fourier transform of six up and down magnetic field sweeps averaged, having performed 3rd order polynomial background subtractions. Here we assume $B = \mu_0 H$. Fourier peaks corresponding to α and β frequencies are seen, as well as a weak feature in the vicinity of 1000T that is seen to be consistent with the presence of a small $2F\alpha$ component when the dominant $F\alpha$ component is subtracted (see supplementary information). **b**, de Haas-van Alphen oscillations at different temperatures, each averaged from six up and down sweeps after background subtraction. **c**, de Haas-van Alphen oscillations of the new frequency $F_\beta = 1654 \pm 40$ T after fitting to and subtracting the contribution from the $F_\alpha = 502 \pm 20$ T frequency oscillations and a small $2\alpha$ plus residual polynomial background (see supplementary information), similarly averaged as in (**b**). The latter fit and subtraction is performed over a shorter interval in field than (**b**) (except for the lowest temperature, see Fig. 3b) to eliminate contamination from flux reversal effects upon reversing the magnetic field sweep direction. The inset in (**a**) is a schematic temperature-doping phase diagram of YBCO. The hole doping is nominally $p_{nom} \approx 0.1$ for $YBa_2Cu_3O_{6.51}$ [ref. 20].

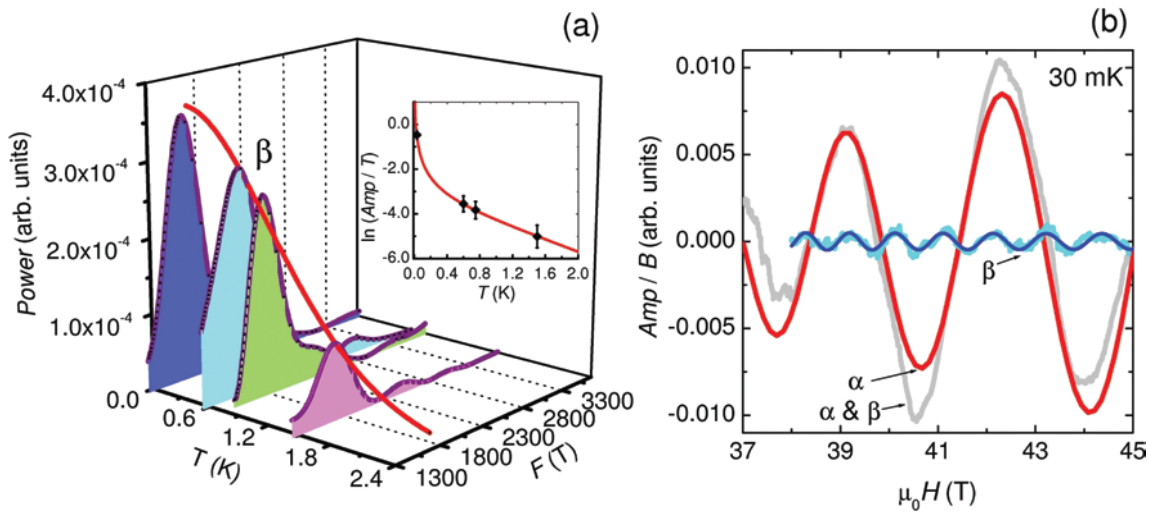

**Fig. 3. Fits to the de Haas-van Alphen oscillations. a**, Power spectrum ($\mathcal{A}^2_T$) of the averaged $\beta$ frequency oscillations in Fig. 2c plotted versus temperature $T$, accompanied by a Lifshitz-Kosevich fit. The inset shows a fit of the amplitude divided by temperature ($\mathcal{A}_T/T$) to the Lifshitz-Kosevich temperature dependence: $\mathcal{A}_T/T = \mathcal{A}_0/\sinh X$, where $X = 14.69\, m^* T/B$ ($m^*$ relative to the free electron mass $m_e$). An effective mass of the $\beta$ pocket is obtained as $m^* = 3.80 \pm 0.4\, m_e$. The functional form of $\mathcal{A}_T$ is derived from the Fermi-Dirac distribution function[11]. **b**, Magnetic field-dependent fits (in red and blue) of $\mathcal{A}_B/B$

$= \mathcal{A}_0 \sin(2\pi F/B+\phi)\exp(-\zeta/B)$ (as would be expected for a two-dimensional Fermi surface)[11] to the measured oscillations (in gray and cyan) of each frequency at $T \approx 30$ mK. The fit including residual background is shown in the supplementary information. For the $\alpha$ pocket, we obtain $\zeta_\alpha = 155 \pm 50$ T, for which possible interpretations in terms of the scattering rate are discussed in the supplementary information. The frequency ratio $F_\beta/F_\alpha \approx 3.29 \pm 0.15$ suggests that the $\beta$ frequency is not an intrinsic harmonic of the $\alpha$ frequency. Since the exponential damping factor scales with the harmonic index, the large value for $\zeta_\alpha$ also implies that the $3F_\alpha$ harmonic of the $\alpha$ frequency should be unobservably small in the Lifshitz-Kosevich theory[11]. Other potential sources of nonlinearities that could lead to harmonics are discussed in the supplementary information – ultimately, a distinction between the $3F_\alpha$ harmonic of the $\alpha$ frequency thus generated and the $\beta$ frequency will depend on the precision to which the frequencies and the relative phase between the oscillations can be measured (see supplementary information). While the $\beta$ oscillations shown in (**b**) are consistent with a fit periodic in $1/\mu_0 H$, an additional field-dependent amplitude prefactor potentially due to the vortex lattice contribution may be indicated by the weaker than expected field-dependence (see supplementary information).

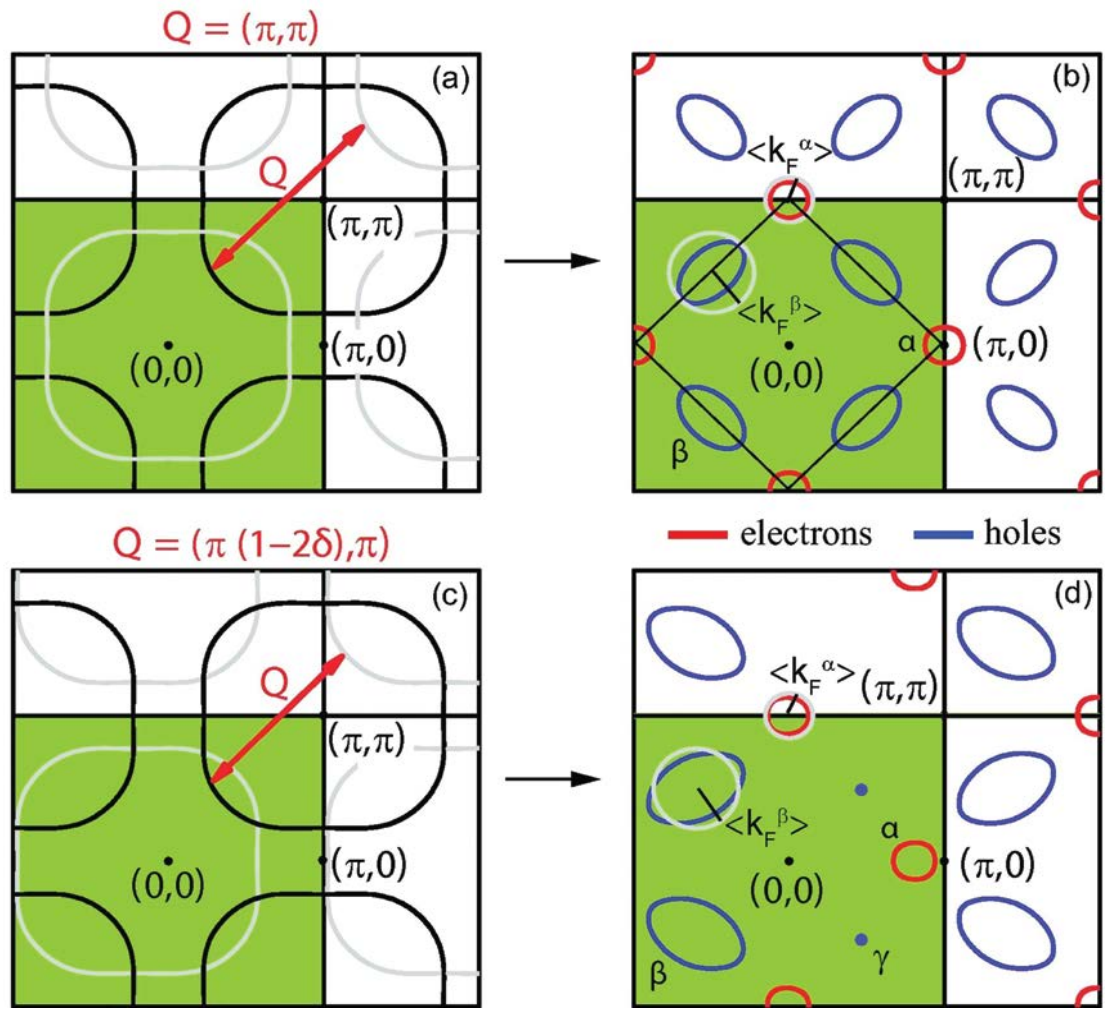

**Fig. 4 Fermi surface reconstruction in YBa$_2$Cu$_3$O$_{6.51}$. a**, Schematic Fermi surface reconstruction for a commensurate ordering wavevector **Q** = ($\pi,\pi$) and nominal doping $p_{nom}$ = 0.1 in the extended Brillouin zone representation. The original unreconstructed Fermi surface (black) represents that for YBa$_2$Cu$_3$O$_{6.51}$ (details in supplementary information). Translation by commensurate **Q** (grey) shifts the original dispersion to $\varepsilon_{k+Q}$, leading to hole pockets at ($\pi/2,\pi/2$), and possibly electron pockets at ($\pi$,0), depending on the magnitude of the gap opening between $\varepsilon_k$ and $\varepsilon_{k+Q}$. **b**, The hole (blue) and electron (red) pockets expected from (**a**) for a nominal doping $p_{nom}$ = 0.1. **c**, Representation of a different translation vector **Q** = ($\pi[1-2\delta],\pi$) assuming $\delta \approx 0.1$, corresponding to a simple helical or spiral spin-density wave scenario (see supplementary information). **d**, Hypothetical Fermi surface on translation by the single **Q** shown in (**c**). Good agreement with the observed pockets is obtained on using $p_{eff}$ = 0.082. For comparison, the respective sizes of the pockets measured experimentally from the average Fermi wavevector $<k_F> = \sqrt{A_k/\pi}$ are represented graphically by grey circles in (**b**) and (**d**). A reduced $\delta$ or ordering gap (or an increased $p_{eff}$) could potentially lead to an additional small hole pocket located at $\gamma$.